\documentclass[12pt]{article}
\usepackage{amsmath,amsfonts, amssymb, braket, bbold}
\usepackage{graphicx,multicol,multirow}
\usepackage{hyperref}
\usepackage{xcolor}
\usepackage{textpos, wrapfig}
\usepackage{titlesec}
\titleformat{\section}
 {\normalfont\fontfamily{put}\fontsize{12pt}{16pt}\bfseries\color{black}}
{\thesection}{1em}{}
\titleformat{\subsection}
 {\normalfont\fontfamily{put}\fontsize{12pt}{16pt}\bfseries\color{black}}
{\thesubsection}{1em}{}
\definecolor{maroon}{rgb}{.45, .2, .05}
\definecolor{medblue}{rgb}{.35, .4, 1}
\definecolor{darkblue}{rgb}{.2, .1, .65}
\definecolor{darkgreen}{rgb}{0, .7, 0}
\definecolor{lightgrey}{rgb}{.85, .90, .9}
\definecolor{brown}{rgb}{.25, .25, .25}
\definecolor{lightblue}{rgb}{.73, .83, .99}
\definecolor{lightbrown}{rgb}{.88, .8, .55}
\definecolor{goldenrod}{rgb}{.80392, .60784, .11373}
\definecolor{darkgoldenrod}{rgb}{.5451, .39608, .03137}
\definecolor{darkolivegreen}{rgb}{.33333, .41961, .18431}
\definecolor{darkred}{rgb}{.75, .15, .15}
\definecolor{cyellow}{rgb}{1.0, 0.812, 0.004} 
\definecolor{cred}{rgb}{1.0, 0.220, 0.224}
\definecolor{mellow}{rgb}{.847, .72, .525}
\definecolor{orange}{rgb}{1.00, 0.65, 0.00}
\definecolor{deepred}{rgb}{.90, .10, .10}
\definecolor{lemonchiffon}{rgb}{1.00, .98, .80}
\def \beq  {\begin{equation}}
\def \eeq  {\end{equation}}
\def \beqar {\begin{eqnarray}}
\def \eeqar {\end{eqnarray}}
\allowdisplaybreaks
\def\sqr#1#2{{\vcenter{\vbox{\hrule height.#2pt
\hbox{\vrule width.#2pt height#1pt \kern#1pt
\vrule width.#2pt}\hrule height.#2pt}}}}

\def\vx {{\vec x}}

\def\vf {{\varphi}}

\def\vp {{\vec p}}

\def\Tr {{\rm Tr}}

\def\vx {{\vec x}}

\def\del {\partial}

\def\A {{\cal A}}

\def\H {{\cal H}}

\def\vf {{\varphi}}

\mathchardef\mhyphen="2D
\begin{document}
\fontfamily{bch}\fontsize{12pt}{17pt}\selectfont
\def \CMP {{Commun. Math. Phys.}}
\def \PRL {{Phys. Rev. Lett.}}
\def \PL {{Phys. Lett.}}
\def \NPBProc {{Nucl. Phys. B (Proc. Suppl.)}}
\def \NP {{Nucl. Phys.}}
\def \RMP {{Rev. Mod. Phys.}}
\def \JGP {{J. Geom. Phys.}}
\def \CQG {{Class. Quant. Grav.}}
\def \MPL {{Mod. Phys. Lett.}}
\def \IJMP {{ Int. J. Mod. Phys.}}
\def \JHEP {{JHEP}}
\def \PR {{Phys. Rev.}}
\def \JMP {{J. Math. Phys.}}
\def \GRG{{Gen. Rel. Grav.}}
\begin{titlepage}
\null\vspace{-62pt} \pagestyle{empty}
\begin{center}
\vspace{1.3truein} {\large\bfseries
~}
\\
{\Large\bfseries ~G\"ursey, Groups \& Fluids ~}\\
\vskip .5in
{\Large\bfseries ~}\\
{\sc V.P. Nair}\\
\vskip .2in
{\sl Physics Department,
City College of New York, CUNY\\
New York, NY 10031}\\
 \vskip .1in
\begin{tabular}{r l}
{\sl E-mail}:&\!\!\!{\fontfamily{cmtt}\fontsize{11pt}{15pt}\selectfont 
 vpnair@ccny.cuny.edu}\\
\end{tabular}
\vskip .5in

\centerline{\large\bf Abstract}
\end{center}
Feza G\"ursey was a brilliant physicist who loved symmetries and beautiful mathematical structures such as group theory and division algebras. In this centenary year, I give a brief review of some of his contributions and then zero-in on group theory and discuss how it can give us a new perspective on a centuries-old topic, namely, fluid dynamics, including new facets such as the chiral magnetic effect.
\vskip .1in
\noindent This is a slightly edited and enhanced version of a colloquium at the
Middle East Technical University on December 9, 2021, which was
organized in celebration of the centenary of Feza G\"ursey. The video 
of the presentation can be accessed by following the link given 
\href{http://www.physics.metu.edu.tr/Department/FezaGursey2021}{here}
or by
clicking
\href{https://www.youtube.com/watch?v=DstVVuOlqGk}{here} for a direct
access.

\end{titlepage}
\pagestyle{plain} \setcounter{page}{2}
\noindent When I think of Feza G\"ursey, always, a few words from a poem by
John Keats come to mind:
\vskip .1in
\noindent
\begin{minipage}{9cm}
\begin{quotation}
\hspace{-24pt}
{Beauty is truth, truth beauty}
\end{quotation}
\flushright{- John Keats, ``Ode on a Grecian Urn"}
\end{minipage}\hskip .3in
\begin{minipage}{5cm}
\scalebox{.8}{\includegraphics{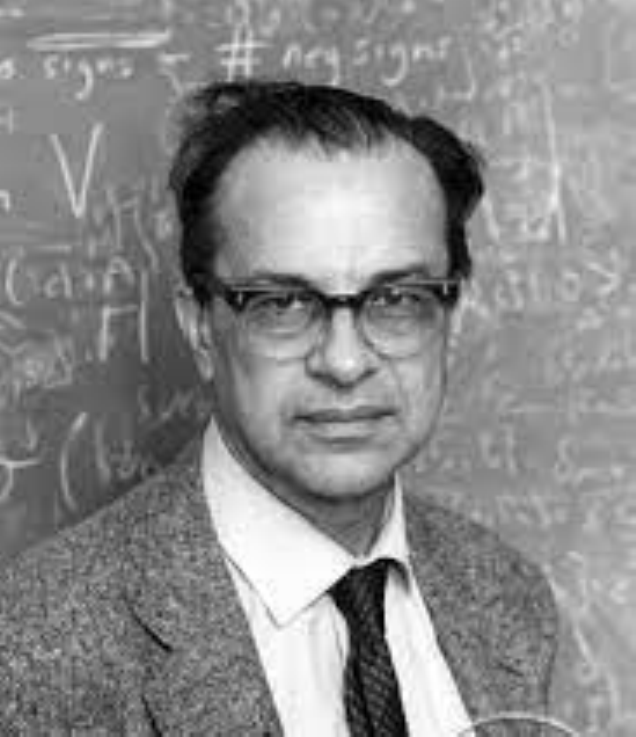}}
\end{minipage}
\vskip .1in
\noindent This is an often-quoted line; some might argue, it is even overused to the point of rendering it boring, but please bear with me a moment for there are people who not only appreciate the deeper meaning of these
words, but who internalize this
and actually live these words. In my view Feza G\"ursey was such a person, a gentleman who impressed everyone with his civility, a brilliant scholar who loved symmetry, groups, quaternions, octonions and other beautiful mathematical structures to the extent of being convinced that, just by virtue of being beautiful,
they should have a bearing on physics. And if anything,
 his own research helped to reaffirm his convictions. 
 
 I was fortunate to have known him in the late 1980s.
I had just joined Columbia University and I would drive up to Yale on some afternoons
for long conversations with him towards the end of his life. Ever the gentleman, sometimes he would kindly invite me to stay for dinner and continue discussions at his house.
In tune with his own love of beautiful mathematics, 
we talked about quaternion analyticity, along the lines of his well-known review article 
with Chia Tze \cite{GTze}.

\begin{figure}[!t]
\begin{center}
\fcolorbox{blue}{lemonchiffon!60}{
\begin{minipage}{15cm}
Choosing real Cartesian coordinates $(t, x)$  in the complex plane,
with $z= t+ i x$, an analytic function $f$ obeys the holomorphicity condition or the Cauchy-Riemann equations
\beq
{\del f \over \del t} +i {\del f \over \del x} = 0
\label{g1}
\eeq
Similarly, a quaternion analytic function can be defined by
\beq
{\del f \over \del t} + i {\del f \over \del x} + j {\del f \over \del y}  + k {\del f \over \del z} = 0
\label{g2}
\eeq
where the 4d-quaternionic plane has coordinates $q = t+ i x +j y +k z$,
$i, j, k$, being the three imaginary quaternions obeying
\beq
 i^2 = j^2 = k^2 = -1, \hskip .3in
 i\, j = k, \hskip .1in j\, k = i, \hskip .1in k\, i = j
 \label{g3}
 \eeq
A complex function which is analytic in a domain $D$ obeys the Cauchy
integral theorem,
\beq
\oint_{\del D} dz\, f(z) = 0
\label{g4}
\eeq
Again, in complete similarity with this,
a function on $\mathbb{R}^4$ which is quaternion-analytic in a domain
$D$
obeys
\beqar
\oint_{\del D} Dq\, f(q) &=& 0\label{g5}\\
Dq &=& dx\wedge dy\wedge dz -i dt\wedge dy\wedge dz
-j dt\wedge dz\wedge dx -k dt\wedge dx\wedge dy
\nonumber
\eeqar
where the contour is now a closed three-surface.  There is also an analogue of the Cauchy integral formula,
\beq
f(z_0) = {1\over 2\pi i} \oint_{\del D} dz\, {f(z) \over z- z_0} ~\longleftrightarrow~
f(q_0) = {1\over 2\pi^2} \oint_{\del D} Dq\, {1\over (q - q_0) \vert q- q_0\vert^2}\, f(q)
\label{g6}
\eeq
For more details, see the review article by A. Sudbery \cite{sud}, as also the
paper by Feza and Chia Tze \cite{GTze}.
\end{minipage}}
\end{center}
\end{figure}

\begin{figure}[!t]
\begin{center}
~\hskip -.7in\scalebox{.85}{\includegraphics{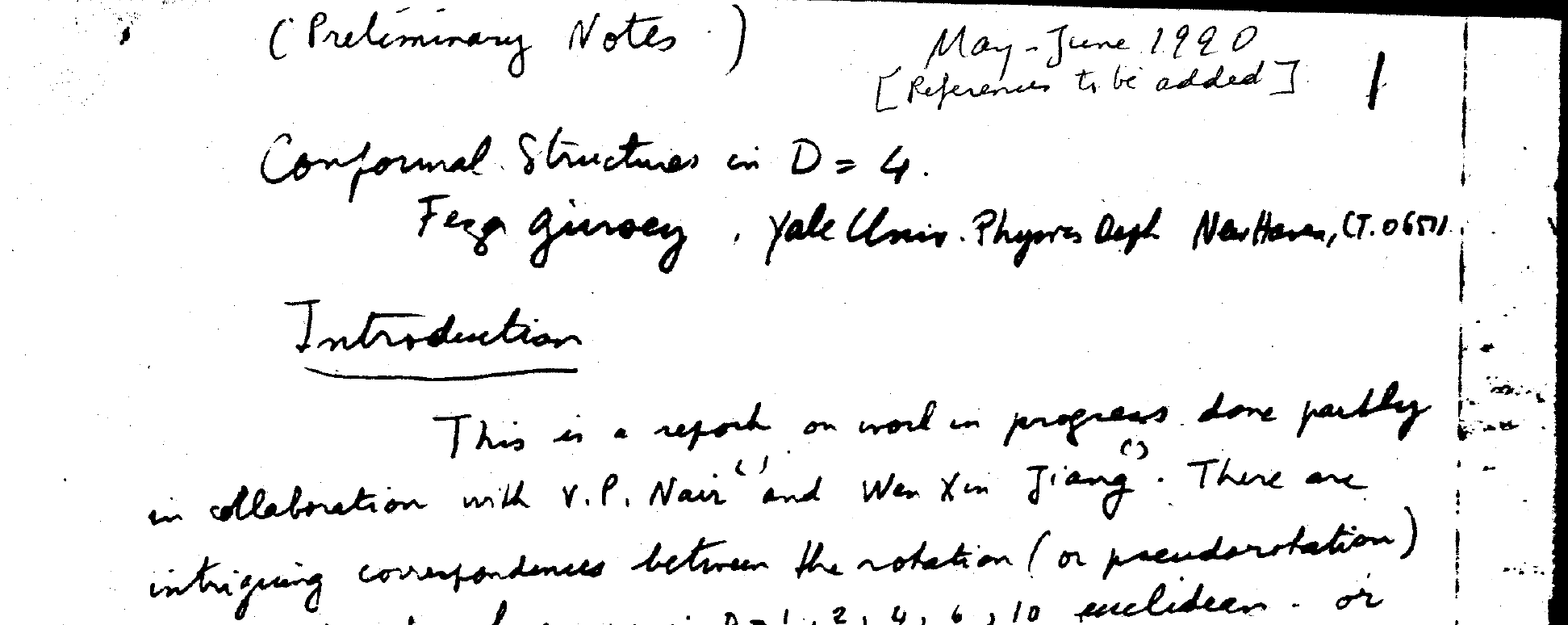}}
\end{center}
\caption{Title part of notes Feza wrote on our discussions, from May-June 1990.}
\label{fig1}.
\end{figure}

Just as in the case of complex variables, one can define analyticity for quaternionic functions. Analogues of Cauchy's theorems
can be proved, with closed three-dimensional surfaces
 (enclosing a four-dimensional volume) taking the place of
closed Cauchy contours (see inset below). We thought this would help us extend notions like the Virasoro algebra to quaternions; it would also involve the conformal
group,
another of Feza's favorites.
Two-dimensional conformal field theory was very much in vogue at that time, both for
applications in its own right and as the underpinning for string theories.
The resolution of correlators in terms of holomorphic
and antiholomorphic conformal blocks
which carried representations of the Virasoro algebra
is at the center of such theories, so seeking a four-dimensional extension
via quaternion analyticity was very natural.
We made some progress, Feza even wrote up some notes
on our discussions, see Fig.\,\ref{fig1}.
But everything
did not quite work out to our satisfaction, perhaps because we aimed too high, seeking a clear and complete result, 
so we did not publish anything, 
but we certainly had a lot of interesting discussions.
Our interests then moved in slightly different directions,
Feza applied similar ideas to
self-dual Yang-Mills fields, he did write a paper
with Mark Evans and V. Ogievetsky, published after his passing
\cite{EGO}.
My student Jeremy Schiff and I also considered (anti)self-dual Yang-Mills fields, but in the context of K\"ahler-Chern-Simons theory, which provides another path to higher dimensions \cite{NS}.
Quaternions, of course, do have a place in many other contexts in physics as well, a notable example being their use in the ADHM (Atiyah-Drinfel'd-Hitchin-Manin) construction of instantons.

Talking about Feza's contributions, perhaps, the most impactful one has been his work on the nonlinear sigma models or more accurately on nonlinear realizations of symmetry.
In the 1950s, physicists were struggling to understand
the strong nuclear forces, especially the pion-nucleon interactions.
The strength of the interaction was too large to allow for any
perturbative analysis, and the weak decays of the pions added another dimension to the difficulties. Key ideas such as $V-A$ interactions and
the partial conservation of the axial current (PCAC) were being discussed and it was in this environment, in 1959, that Feza wrote his paper 
``On the Symmetries of Strong and Weak Interactions" \cite{Gur1}.
(I have displayed a few equations from this paper as Fig.\,\ref{fig2}).
In it he considered a model where one could carry out independent 
isospin ($SU(2)$) transformations (with parameters
${\boldsymbol\mu}$ and ${\boldsymbol\nu}$) on the left and right chiral components
of the nucleon (the isospin doublet made of the proton and the neutron).
The pion fields $\vf^a$
(there are three such fields corresponding to $\pi^{\pm} \sim
(\vf^1 \pm i \vf^2), \pi^0 \sim \vf^3$)
are realized
as the parameters of the axial transformations, 
they are the angular variables in terms of a unitary matrix $\Phi$.
Feza refers to the full symmetry as $G_4$, it
corresponds to $SU(2)\times SU(2)$ in present-day terminology.
The second equation gives the transformation of the pion fields.
If one works out ${\boldsymbol\vf}'$ in terms of
${\boldsymbol\vf}$ and the parameters
${\boldsymbol\mu}$, ${\boldsymbol\nu}$, using the 
Baker-Campbell-Hausdorff
formula for combining exponents,
it will be a rather long and involved expression which is nonlinear in
${\boldsymbol\vf}$. This is the origin of the terminology of nonlinear
realization of symmetry. The full action for the theory
will have the symmetry under both the left and right
tranformations.
The third equation in Fig.\,{\ref{fig2} gives the
pion-nucleon interaction consistent with the
nonlinearly realized $SU(2) \times SU(2)$ symmetry.
Upon expanding the exponential, we find the mass term
and the pion-nucleon coupling $i g \gamma_5 {\boldsymbol\tau}
\cdot {\boldsymbol\vf} \psi$ as the two lowest order terms.
\begin{figure}[!t]
\begin{center}
\scalebox{1}{\includegraphics{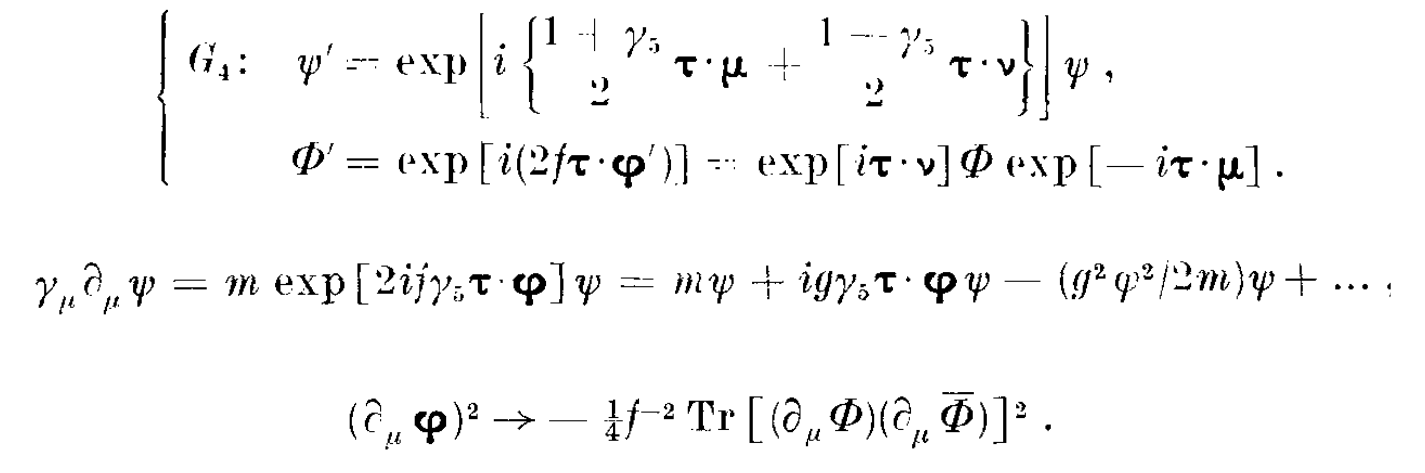}}\\
\end{center}
\null\vspace{-36pt}
\caption{Some of the key equations from F. G\"ursey,
``On the Symmetries of Strong and Weak Interactions" \cite{Gur1}. }
\label{fig2}
\end{figure}
Remarkably, this coupling constant $g = 2 f\, m$ is also the lowest order version
of the famous Goldberger-Treiman relation \cite{{GT},{currenta}}.
(It would be the correct one at the level of point-particles, 
which is the approximation Feza used in his model.
The constant $f$, sometimes denoted by
$f_\pi$, is the pion decay constant which controls the lifetime
of the charge pions due to weak decays.
Today, with our understanding of nucleons as composite particles
made of quarks, we can see that this relation will be modified
due to nonpoint-like nature of nucleons, i.e., due to
form factor effects. This changes the coupling to $g = g_A \, 2 f\, m$, with
$g_A \neq 1$. The actual observed value of $g_A$ is approximately
1.2; this was calculated using a sum rule obtained via
current algebra methods by Adler and
Weisberger, with numerical estimate obtained by using
experimental data on scattering \cite{{AW},{currenta}}.)
Finally, the last equation is the effective Lagrangian for the 
pions in the nonlinear
realization of the symmetry.

Although Feza had proposed the idea of nonlinear realization
of symmetry in 1959, it took a
while to catch on. A few months after Feza's work, Gell-Mann and Levy
proposed a linear version where the symmetries are realized
linearly on the bosonic fields as well \cite{G-ML}. This requires an additional field
which they designated by the Greek letter $\sigma$.
This paper attracted a lot of attention and led to the name
$\sigma$-model for this type of theory, so the nonlinear realization is
sometimes referred to as the nonlinear $\sigma$-model, although
there is actually no $\sigma$ in it.

In the early 1960s, attempts
to understand strong interactions and most of early results were based
on the technique of current algebra. 
Pion-nucleon scattering was calculated in this way by
many people: Steven Weinberg \cite{Wein1}; Balachandran, Gundzik,
Nicodemi \cite{BGN}; Tomozawa \cite{Tom}; Raman and Sudarshan
\cite{RSud}; and others.
Around
1966-'67, Steven Weinberg was able to express
results on pion-nucleon interactions, obtained via current algebra
techniques,
in terms of an effective action \cite{Wein2}.
The resulting action agreed 
with the nonlinear
version of the sigma model which had been proposed earlier by Feza,
not the linear version.
Although Feza did not include a mass term for the pions in
his 1959 paper, it can be easily included. This
``PCAC-modified" version agrees with the results obtained by
Weinberg, as pointed out by Chang and G\"ursey \cite{CGur}.
Weinberg was, in fact, rather critical of the Gell-Mann-Levy
linear model, since the $\sigma$-field does vitiate the
current algebra results due to additional
interactions.
It is interesting to note that he makes the remark
``{\it The $\sigma$ model is also useless as a phenomological
model of strong interactions;}"\footnote{I quote exactly, including the
typo from the original paper \cite{Wein2}.}

Today we have a much better understanding of strong interactions
in terms of Quantum Chromodynamics or QCD, which is a nice renormalizable theory. But to date,
the equations in Feza's 1959 paper (with some modifications which are conceptually minor)
constitute the effective description of low energy
dynamics in QCD. Thus it has become an integral part of the Standard Model of particle physics.

The method of nonlinear realizations of symmetry
is not limited to high energy physics. It has
become the standard for the effective description 
of spontaneous symmetry breaking, which is 
ubiquitous in modern physics, whether we talk about pseudoscalar
mesons like the pion, or spin waves in a ferromagnet,
or phonons in a crystal lattice, and so on.
If a continuous symmetry with group $G$ is spontaneously broken to
a subgroup $H$, then there are gapless excitations (i.e., massless 
excitations in the relativistic case), the Goldstone bosons,
one field for every broken symmetry. The dynamics of these fields at
low energies is given by an action of the form
\beq
S = {f^2 \over 2} \int d^4x~ g^{\mu\nu} G_{ab} (\vf) \, \del_\mu\vf^a 
\del_\nu \vf^b
\label{g7}
\eeq
where $\vf^a$ is a map from spacetime to $G/H$. Thus they are like the coordinates on the space $G/H$ and $G_{ab}$ is the metric
tensor (which can be $\vf$-dependent) on this space. This is a universal result based purely on symmetry arguments. The symmetry corresponding to $H$ is linearly realized, with the remaining ones realized nonlinearly.
If one takes the case $G = SU(2)\times SU(2)$ and $H = SU(2)$,
this action (\ref{g7}) becomes the action for pions
in the last line of Fig.\,{\ref{fig2}. We see that the action Feza wrote down
in 1959 is the prototype for spontaneous symmetry breaking.

In 1964, G\"ursey and Radicati proposed the
$SU(6)$ symmetry for baryons and mesons \cite{GRad}. Independently,
Bunji Sakita also proposed the same model \cite{Sak}. This symmetry combines spin and flavor symmetry for hadrons, basically a generalization of
Wigner's $SU(4)$ symmetry in nuclear physics.
Since spin is part of the same group $SU(6)$, this model can lead to
mass formulae
connecting particles of different spin, e.g.,
\beq
m_{K^*}^2 - m_\rho^2 = m_K^2 - m_\pi^2
\label{g8}
\eeq
relating the masses of $K^*$ and $\rho$ which have spin one to
those for the $K$- and $\pi$-mesons.
Needless to say, the relation is reasonably good, within the limits of expected
deviations for such analyses. The $SU(6)$ symmetry also
leads to relations connecting magnetic moments
of the hadrons.

In the 1975, G\"ursey, Ramond and Sikivie proposed the grand unified theory based on the exceptional group $E_6$
\cite{GRS}, providing another path to grand unification
beyond the $SU(5)$ and $SO(10)$ models.
Exceptional groups, of course, were a most natural province for a man in love with quaternions and octonions. 
The decay of the proton would be the most telling signature for
grand unification since it combines the strong and electroweak theories
within a simple group. We have not seen proton decay, despite searches lasting decades, so it is fair to say that
we do not yet have any experimental evidence for any kind of grand unification, let alone which specific group might be relevant.
Nevertheless, if we envisage obtaining the Standard Model
from a larger construct such as, for example, string theory,
it could very well be that the cascade of symmetry breakings needed to
get to the Standard Model runs through $E_6$, especially since
$E_8$ has a natural place in some string theories.
In short, we could say the jury is still out on this matter.

In the 1980s, Feza,
with Sultan Catto, also generalized the $SU(6)$ symmetry further
to connect bosons and fermions using supergroups (e.g., $SU(6|21)$)
\cite{GCat}.
Similar supersymmetry had been used as a classifying scheme
in nuclear physics by Miyazawa and by Iachello
\cite{Iach}.
While $SU(6)$ can give formulae connecting
 masses and magnetic moments for particles differing by integer spin,
 with supersymmetry, one can get additional results connecting 
 integer and half-integer spins, i.e., bosons and fermions.
 
 Feza has a large number of other publications which I
 cannot hope to review in any detail here.
These include several papers on group theory applied to
scattering, work on the conformal group, on noncompact groups,
on certain interesting gravitational backgrounds, 
even some noncommutative geometry.
It is clear that group
theory occupied a large part of Feza's professional life.
 Perhaps, the best commendation for his lifelong engagement
 with group theory is the Wigner Medal with which he was honored in 1986.
 
 I am of the opinion that in paying tribute to physicists, 
 personal reminiscences and evaluation of their contributions
 are important and do have a place, but  it is also important
 to look beyond to something new, perhaps something new we can say
 following the general philosophical attitude they had to how they
 conducted
 their own research.
Looking through Feza G\"ursey's papers, we find a number of strikingly original ideas, we see that even when he 
contemplates old problems, there are
new and useful insights. 
So in appreciation, and considering his love of group theory,
I would like to pick a topic, specifically fluid dynamics,
and follow his philosophy of applying group theory to it to see what
new insights we might get.
 
Fluid dynamics, as you know,  is a topic
as old as
physics itself, going back to the ancient Greeks, perhaps earlier, maybe to ancient Mesopotamia, since channeling and modulating
water flow has been a practical necessity for the development of
human settlements.
Even the modern scientific formulations go back a few hundred years,
 to Torricelli, Newton, Bernoulli, Euler,
Lagrange and others. 
With so many physicists having worked on it for so many years,
one might wonder if there is really anything new one
can say on this matter. I hope to convince you that there are still many interesting things to say, many interesting extensions possible
\cite{Fl-rev}.

To set the stage, let me start by remembering how Lagrange approached the problem of fluids. Unlike Euler who described fluids directly in terms of density and a local velocity, Lagrange started by thinking of fluids as made of a large number of point-particles.
Each particle may be taken to obey Newton's laws of motion. We can express them as the well-known equations
\beq
m\, {d^2 X^i (\alpha) \over dt^2} = - {\del V \over \del X^i(\alpha)},
\hskip .3in i = 1,2, 3; \hskip .1in \alpha = 1, 2, \cdots , N
\label{gur1}
\eeq
Here $\alpha$ labels the particle and ${\vec X}(\alpha) =
(X^1(\alpha), X^2(\alpha ), X^3(\alpha))$ denotes the position of the 
$\alpha$-th particle.
As time moves forward, say from zero to $t$, the particles will move from a set of initial positions $X^i(\alpha, 0)$ to a set of final
positions $X^i(\alpha, t)$. Lagrange's key insight was to realize that one could label the particles by the initial positions, so that we could write
\beq
X^i(\alpha, 0) = \alpha^i
\label{gur2}
\eeq
Classically, time-evolution is a nice continuous process, so we have a mapping of positions $\alpha^i \rightarrow X^i(\alpha, t)$, which is also a nice
smooth map. In the limit of a large number of particles, 
there is a particle at every point $(\alpha^1, \alpha^2, \alpha^3)$, if
we look at the distribution of particles 
in a coarse-grained sense, with a resolution coarse enough to regard a small
volume around each particle as a point.
The map $\alpha^i \rightarrow X^i(\alpha, t)$ is thus a
smooth map of space into
itself. More precisely, we have a time-dependent diffeomorphism
$X^i (\alpha, t)$.
 \begin{figure}[!t]
 \begin{center}
\scalebox{1}{\includegraphics{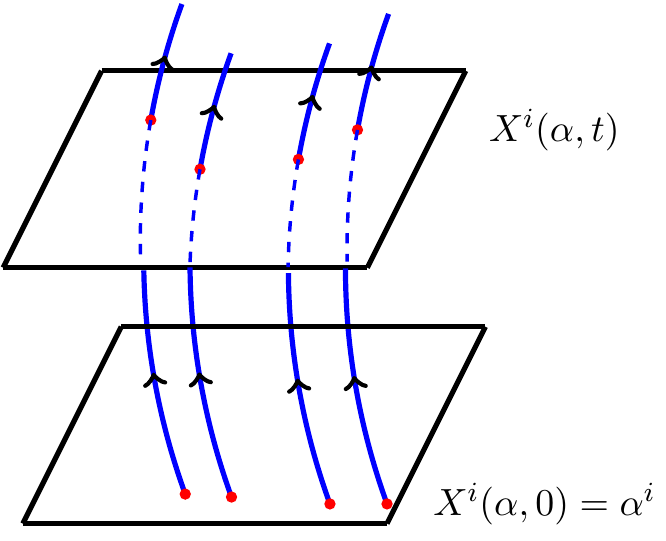} }
\end{center}
\caption{Showing how particle trajectories give a smooth map from
the spatial manifold viewed as initial data (at time $t = 0$) to
the spatial manifold viewed as particle positions at time $t$.}
\label{fig3}
\end{figure}
It should be possible to describe fluids using this idea. For example,
since particles do not just appear spontaneously or vanish, we have the conservation equation
\beq
\int \rho^{(0)}  (\alpha, 0 ) \, d^3 \alpha = \int \rho (X, t) \, d^3X
\label{gur3}
\eeq
We can then define a density $\rho (X, t)$ by
\beq
\rho (X, t) \, \bigg\vert {\del X \over \del \alpha} \bigg\vert
= \rho^{(0)} (\alpha, 0)
\label{gur4}
\eeq
where $\vert \del X/\del \alpha\vert$ is the Jacobian of the
transformation of coordinates $\alpha \rightarrow X$.
Since $\rho (\alpha, 0)$ is independent of time, just
by direct differentiation, it is easy to check that $\rho(X, t)$ obeys the identity
\beq
{\del \rho \over \del t} + \nabla \cdot (\rho {\vec v}) = 0,  \hskip .3in
{\vec v} = {\del {\vec X} \over \del t}
\label{gur5}
\eeq
We see that we get the familiar equation of continuity for fluids.
In a similar way we can get the Euler equation from Newton's law
(\ref{gur1}) by the same coarse-graining procedure.

It is not so much the mathematical derivation which I want to focus on, but the key idea: Start with point-particles, take a continuous (or coarse-grained) distribution of such particles and work out the equations.
This is Lagrange's trick for getting the equations for a fluid. It is such a simple and beautiful idea and it begins with the notion of a point-particle.
But today, more than 200 years after Lagrange, when we think about it, we are already stuck at the very beginning: {\it What exactly is a point-particle?}
Classically, we could think of a point-particle as the idealization of a small ball of almost zero radius. But today, with our understanding of quantum mechanics, this is no longer acceptable.
In modern physics, a point-particle must be defined as a unitary irreducible representation of the Poincar\'e group. These are complicated words and I will explain them in a minute.
The point is that we must only deal with observable quantities and these are the position $\vx$ and momenta $\vp$ for a particle. Quantum mechanically,
these obey the famous rules
\beqar
{\hat x}^i {\hat x}^j - {\hat x}^j {\hat x}^i &=& 0\nonumber\\
{\hat x}^i {\hat p}_j - {\hat p}_j {\hat x}^i &=& i \hbar\, \delta^i_j\label{gur6}\\
{\hat p}_i {\hat p}_j - {\hat p}_j {\hat p}_i &=& 0
\nonumber
\eeqar
and we must have a realization (or representation) of these, maybe in the familiar form
\beq
{\hat x}^i \, \psi = x^i \, \psi , \hskip .3in
{\hat p}_i \, \psi = - i \hbar {\del \over \del x^i} \psi
\label{gur7}
\eeq
Basically, this, a realization of the basic algebra (\ref{gur6}) (which is
 referred to as the Heisenberg algebra), defines a particle. 
The Hamiltonian, which can be written as an operator
in the representation (\ref{gur7}), will depend on which physical system
we are looking at, whether it is a free particle, a particle bound to some attractive center (like the electron in the Hydrogen atom), etc.
 
Once we include relativity, things are a little more involved. Energy and momentum should go together as a four-vector
$p_\mu = (E, p_i)$ and we also have Lorentz transformations and rotations,
generated by $J_{\mu\nu} $. The enhanced algebra is then
\beqar
{\hat p}_\mu \, {\hat p}_\nu -  {\hat p}_\nu {\hat p}_\mu &=& 0\nonumber\\
{\hat J}_{\mu\nu} {\hat p}_\alpha - {\hat p}_\alpha  {\hat J}_{\mu\nu} &=& i\,(g_{\mu\alpha} {\hat p}_\nu -
g_{\alpha \nu} {\hat p}_{\mu} ) \label{gur8}\\
{\hat J}_{\mu\nu} {\hat J}_{\alpha\beta} 
- {\hat J}_{\alpha\beta}  {\hat J}_{\mu\nu} &=& i\left(
g_{\mu\alpha}{\hat J}_{\nu\beta} ~+~g_{\nu\beta}{\hat J}_{\mu\alpha}~-~
g_{\mu\beta}{\hat J}_{\nu\alpha}~-~g_{\nu\alpha}{\hat J}_{\mu\beta}\right)\nonumber
\eeqar
These relations define what is called the Poincar\'e group.
So a realization (or representation)
of these relations on some wave function should be used as the definition of a relativistic point-particle.
The particle will be characterized by mass and spin, with possible states labeled by momentum and angular momentum.
(The representation should be unitary to preserve probabilities in the quantum mechanical sense. It should be irreducible because we are considering only one particle, not a collection of particles with different spins and masses.)
This way of thinking about particles
brings us back to group theory, for this
definition is in terms of groups, the Poincar\'e group to be specific.
Ever since Wigner's work on representations of the Poincar\'e group in 1939, we have known that this is the proper way to talk about point-particles, namely as
a unitary irreducible representation of the Poincar\'e group.
Admittedly, this is rather abstract but has the advantage that it does not speak of anything that is not observable.
So let us pose the question again: {\it With this more modern definition, can we do Lagrange's
trick of a continuous or coarse-grained distribution and obtain a new
avatar of fluid dynamics?}
The answer is, of course, yes, but there is one subtlety we must take account of.
Classically, one can deal with the equations of motion, e.g., Newton's laws,
but here we are moving towards a quantum definition of a particle and so
what is a needed is a Lagrangian (or a Hamiltonian which can then be used to
formulate a Schr\"odinger equation).

This brings us back again, closer to the work Feza has done. The key element in his work on nonlinear realization of symmetry is an action of the form
\beq
S = - {1\over 2} \int dt \,\Tr \left( g^{-1} {\dot g} \right)^2, \hskip .3in
{\dot g} = {d g \over d t}
\label{gur9}
\eeq
where $g$ is an element of a group $G$, presented here as
a matrix.\footnote{I now use the letter $g$ for a group element instead of
$\Phi$ which was used in Feza's 1959 paper. Also I do not include spatial
coordinates for now, will bring them in later.}
 If we use a rotation matrix for $g$, this is like the action for a rigid rotor.
 The Hamiltonian for the action (\ref{gur9}) will be exactly of the
 form $J^2 = {\vec J}\cdot {\vec J}$, where ${\vec J}$ denotes the angular momentum operator.
 If we solve
the Schr\"odinger equation for this case, we will
obtain states with all possible
angular momenta. From a group theoretic point of view, we get
all possible (unitary irreducible) representations.
This is a bit too much, what we need for a particle is just one unitary irreducible representation
of the Poincar\'e group. So the next natural question is: {\it Is there an action, may be some simpler version of (\ref{gur9}), we can write down, which leads,
in the quantum theory, to exactly one unitary irreducible representation of a group?}
There is indeed such an action. It emerged from the work of Borel, Weil and Bott in the 1950s and from later work on geometric
quantization by Kostant, Souriau and Kirillov
on the mathematical side \cite{geom}, and Wong \cite{Wong} and Balachandran {\it et al} 
\cite{Bal+} on the physics side.
This action is given by
 \beq
 S = i  \sum_a w_a \, \int dt ~ \Tr ( h_a \, g^{-1} {\dot g} ),
 \hskip .3in {\dot g} = {d g \over d t}
 \label{gur10}
 \eeq
 where $h_a$ give a basis of the diagonal generators of the Lie algebra (the Cartan subalgebra)
 and $w_a$ are a set of numbers. We are considering $g$
 presented as a matrix, say, in the fundamental representation, with $\Tr (h_a \, h_b ) = \delta_{a b}$.
 The basic theorem is that the quantization of this action leads to a Hilbert space
 which carries a unitary irreducible representation (UIR) of the group $G$, this UIR being specified by the highest weight $(w_1, w_2, \cdots, w_r)$, $r$ being the rank of the group, which is also the range of summation for
 $a$. The canonical one-form associated to (\ref{gur10}) is evidently
 \beq
 \A = i \sum_a w_a \, \Tr ( h_a \, g^{-1} { d g} )
 \label{gur11}
 \eeq
 Under transformations $g \rightarrow g \, \exp (-i h _a \vf_a )$, we find
 $\A \rightarrow \A + d f $, $f = \sum w_a \vf_a$.
 Thus the symplectic two-form $\Omega = d \A$ is defined on $G/T$, $T$ being the maximal torus.
 Further, the transformation $\A \rightarrow \A + df$ shows that in the quantum theory, where wave functions transform as $e^{iS}$, there will be restrictions or quantization conditions on $w_a$,
 these being the appropriate conditions for $(w_1, w_2, \cdots, w_r)$ to qualify as the highest weight
 of a UIR.
 These numbers, $(w_1, w_2, \cdots, w_r)$,
 will thus determine the characteristics of the particle.
 
 We can easily generalize the action (\ref{gur10}) to a collection of 
 particles indexed by $\alpha$ as
 \beqar
 S &=& i  \sum_\alpha \sum_a w_{\alpha a}  \, \int dt ~ \Tr ( h_a \, g_\alpha^{-1} {\dot g}_\alpha )\nonumber\\
 &\rightarrow& i \sum_a \int d^3\alpha\, dt\,  \rho^{(0)}_{a} (\alpha) 
 \int dt ~ \Tr ( h_a \, g_\alpha^{-1} {\dot g}_\alpha )\nonumber\\
 &=& i \sum_a \int d^4x\, \rho_a (x)\, 
 \Tr ( h_a \, g^{-1} {\dot g} )
 \label{gur12}
 \eeqar
 In the second line, we use Lagrange's trick of treating $\alpha^i$ as
 continuous variables in the coarse-grained sense, in the third line we go to the coordinate $x^i$ from $\alpha^i$.
 Including a Hamiltonian density $\H$ as well, we may thus write the action we need as
 \beq
 S = i \sum_a \int d^4x\, \rho_a (x)\, 
 \Tr ( h_a \, g^{-1} {\dot g} ) - \int d^4x\, \H
 \label{gur12a}
 \eeq
 Basically, this should give us our version of fluid dynamics, for any kind
 of fluid. (This approach to fluid dynamics is 
 proposed, and different aspects of it are analyzed, in several papers,
 see \cite{Fl-rev}, \cite{Bistro}-\cite{CNT}.)
 
 Notice that we need as many independent densities $\rho_a$ as there are diagonal generators for the group. We may view each independent density
 as the time-component of some current $J^\mu_a$, since it comes naturally multiplied by the
 time-derivative of $g$. So we can generalize (\ref{gur12a})  to a relativistic form
 and write\footnote{I have also included an arbitrary spacetime metric
 $g_{\mu\nu}$ so this equation is applicable even in curved space.}
 \beq
  S =  \int d^4x \sqrt{-\det (g_{\alpha\beta})}\, \left[ i \sum_a  J^\mu_a (x)\, 
 \Tr ( h_a \, g^{-1} {\del_\mu g} ) - F(n)\right]
 \label{gur13}
 \eeq
 where $F(n)$ is some function which is specific to the fluid we are
 considering. Here, for each value of $a$, we define an invariant density $n_a$ by $n_a^2 = g_{\mu\nu} J^\mu_a J^\nu_a$.
 The energy density of the fluid, the pressure and the equation of state
 are encoded in $F(n)$.
Equation (\ref{gur13}) is in a manifestly relativistic
form. 
 The equation of motion for $J^i_a$ is given by
 \beq
 J^i_a = - i \,{n \over F'} \Tr ( h_a g^{-1} \del_i g)
 \label{gur14}
 \eeq
 This shows that the flow velocity is of the form
 $-i\Tr ( h_a g^{-1} \del_i g)$. Eliminating $J^i_a$ by using
 (\ref{gur14}) brings the action (\ref{gur13}) to a form in terms of densities and flow velocities as in (\ref{gur12a}).
 (In (\ref{gur12a}), $\H$ is generally a function of the flow velocities.)

 {\it Essentially all we need for fluid dynamics is
 the action (\ref{gur13}); it should be universally applicable,
 able to describe the dynamics of any fluid}.
It is quite remarkable how, in
a very simple way, just using ideas of group theory,
we have managed to obtain a new, general and modern description of fluids.
We have only used two key ideas: the action (\ref{gur10}) to obtain
a representation of the group and the coarse-graining 
technique of Lagrange.

Let us consider particles with no charge or other internal quantum numbers.
Then the group we should use is just the Poincar\'e group \cite{KarN}.
The diagonal elements correspond to mass and spin.\footnote{Strictly speaking, because the trace is not well defined for the
Poincar\'e group, we should consider the de Sitter group
for which there are two independent diagonal elements for the algebra.
A contraction of this algebra then leads to the results for the
Poincar\'e case, see \cite{KarN}.}
Particles are thus characterized by their mass and spin and
the action (\ref{gur13}) will give flow equations for transport of mass and for transport of spin,
with independent flow velocities for each. It is easy to see that, generically, we should have independent velocities. For example, in
a fluid made of particles with spin, we could consider a small cluster
of particles where the spins of the individual particles combine to a state of total spin equal to zero. If this composite, or a collection of such composites,
moves in some direction, we will get flow of mass or energy
with no transport of spin, see Fig.\,{\ref{fig4}.
This clearly shows that generically we should have independent spin and mass flow, in agreement with the two independent diagonal elements.

\begin{figure}[!t]
\begin{center}
\scalebox{1}{\includegraphics{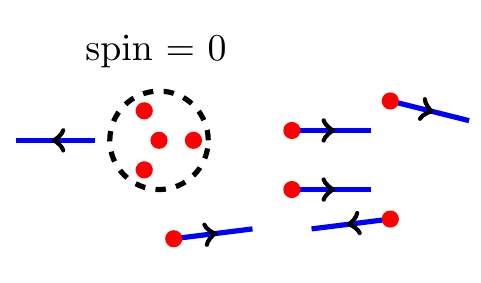}}
\end{center}
\caption{Showing the possibility of independent transport of mass and spin}
\label{fig4}
\end{figure}

If we seek a fluid description of the quark-gluon plasma, we will have
particles which carry color degrees of freedom, corresponding to the
symmetry group $G = SU(3)$ of QCD. The action (\ref{gur13}) for $g \in SU(3)$ can thus describe the dynamics of the quark-gluon plasma \cite{Bistro}. Of course, in this case too,
we can have transport of mass or spin without the corresponding color transport, since there could be local composites which are color singlets.
So the full fluid dynamics of the quark-gluon plasma will be given by
an action of the form (\ref{gur13}) with $G = {\rm Poincar\acute{e}} \times SU(3)$. There will be two currents, corresponding to the flow of
the color isospin (the third component, say) and color hypercharge in the color $SU(3)$, in addition to the flows for mass and spin.

We can go further by considering the full group
$G = {\rm Poincar\acute{e}} \times SU(3) \times SU(2)\times U(1)$
of the Standard Model. A full fluid phase of all the particles relevant to the
Standard Model can be obtained at very high temperatures, say,
above $T\sim 250 \, GeV$. Presumably, such a phase existed
at some point in the very early universe and can be described by the action
(\ref{gur13}) with this choice of $G$ \cite{{NRR},{CNT}}.

So clearly, through the lens of group theory, we have obtained a new perspective 
on fluid dynamics. Shall we now dare to ask if there are new effects,
not just a new point of view, that
we can easily understand in this way? Indeed there are, I shall
mention a few where qualitatively new features are highlighted.
The first is about the minimal case of particles with mass and spin, with no
other quantum numbers of interest, so that the relevant group
is just the Poincar\'e group.
 We will even consider a special case
where spin and mass transport occur with the same flow velocity;
this will suffice to illustrate our point.
Now, one of the hallmarks of relativity for particle dynamics
 is the spin-orbit coupling, enforced mathematically by the fact that
 the commutator of
 two Lorentz velocity boosts, say $K_1 = J_{01}$  and
 $K_2 = J_{02}$, leads to a rotation, $[ K_1, K_2] = i J_{12}$.
 As a result, the build-up of accelerations in different directions (which would generate orbital angular momentum) also leads to spin precessions.
 In a fluid, we should expect that this can lead to spin precession effects
 due to pressure gradients (which can cause local accelerations).
 Indeed, if we take $G = {\rm Poincar\acute{e}}$, the equation for
 spin flow is given by
 \beq
 F'\, (u^\alpha \del_\alpha S^{\mu\nu}) = S^\lambda_\mu (u_\lambda \del_\nu
F' - u_\nu \del_\lambda F') - (\mu \longleftrightarrow \nu), \hskip .2in
F' = {\del F \over \del n}
\label{gur15}
\eeq
where $u^\alpha = (1, v^i)/\sqrt{1- v^2}$ is the 4-velocity of the
spin (and mass) flow and 
$S^{\mu\nu}$ is the spin density.
Since $F(n)$ characterizes the pressure and energy density, we see that we do have a spin-orbit effect at the level of fluid flow \cite{KarN}.
In principle, this equation should be derivable from
standard field theory (although I know of no such derivation), but
symmetry arguments lead directly to (\ref{gur15}) in a very economical
way, and
without the limitations
of perturbation theory which we often need to use for
field theory derivations.

We know that individual test particles in a gravitational background
follow geodesics.
The motion of the fluid, which is not exactly along geodesics
since there is the extra non-gravitational
force term due to pressure gradients, is captured by a conservation equation $\nabla_\mu T^{(f)\mu\nu} = 0$, where
$T^{(f)\mu\nu} $ has the perfect fluid form
\beq
T^{(f)\mu\nu}  = n F' u^\mu u^\nu - g^{\mu\nu} (n F' -F)
\label{gur16}
\eeq
However, we also know that spinning particles do not follow 
geodesic motion.
Rather they obey the
so-called Mathisson-Papapetrou equation where
there is an additional force term involving the direct coupling of the curvature to spin \cite{MathP}. It is again remarkable that the flow equation which we can obtain from (\ref{gur13}) is \cite{NRR}
\beq
\nabla_\mu T^{(f) 
\mu\nu} = 2\, (R_{\alpha\beta})^\nu_{~\lambda} \, j^\lambda S^{\alpha\beta} \label{gur17}
\eeq
where
\beq
j_\mu = -i {n \over F'} \Tr \left( \Sigma_{12} \Lambda^{-1} D_\mu \Lambda
\right), \hskip .3in
S^{\alpha \beta}= \Tr (\Sigma_{12} \Lambda^{-1} \Sigma^{ab} \Lambda )
(e^{-1})^\alpha_{\, a} (e^{-1})^\beta_{\, b}
\label{gur17a}
\eeq
Here $\Sigma^{ab} = [\gamma^a , \gamma^b]/(4i)$ in terms of the
Dirac $\gamma$-matrices, so that $\Sigma_{12}$ is the usual
diagonal spin matrix. Therefore $j_\mu$ in (\ref{gur17a}) denotes the spin flow
current as in (\ref{gur14}) with $g = \Lambda$, an element of the
Lorentz group (in the usual spinorial representation).
$(e^{-1})^\alpha_{\, a}$ is the inverse of the tetrad field, defined by
$g^{\alpha\beta} = \eta^{ab}(e^{-1})^\alpha_{\, a}(e^{-1})^\beta_{\, b}$.
Equation (\ref{gur17}) thus gives the fluid version
of the Mathisson-Papapetrou equation and shows that one can have a spin separation effect in a fluid due to the spacetime curvature.

Since the group theoretic approach is built on symmetry arguments,
it is basically tailor-made for another fascinating phenomenon, the chiral magnetic effect. As is well known, the axial $U(1)$ current $J^\mu_A$ in QCD is not
conserved, but obeys an equation of the form
\beq
\del_\mu J^\mu_A = {e^2\over 2 \pi^2} {\vec E} \cdot {\vec B} + \cdots
\label{gur18}
\eeq
where $\vec E$ and $\vec B$ denote the electric and magnetic fields.
(There are additional terms, denoted by the ellipsis in
(\ref{gur18}), which will not be important for the effect we focus on.)
This equation is a statement of the chiral anomaly.
Since this is a symmetry effect, or rather the expression of a symmetry which
is broken in a very specific way, we can ask if we can represent this in the fluid language. Indeed, it is easily accommodated by adding
an extra term to the action (\ref{gur13}) of the form\footnote{Here we are focusing on the anomaly in just one type of current. It is possible to write down the extra action we need to reproduce all the possible anomalies of the Standard Model in terms of the fluid variables, see \cite{{NRR},{CNT}}.
For another case of $U(1)$ anomaly in fluids, see \cite{MAN}.}
\beq
S_{\rm extra} =  {e^2 \over 4 \pi^2}  \int \epsilon^{\mu\nu\alpha\beta} 
A_\mu\, V_\nu \, \del_\alpha A_\beta
\label{gur19}
\eeq
where $A_\mu$ is the electromagnetic vector potential
and $V_\nu = \del_\nu \vf$ is related to the flow velocity for
the axial charge. (Notice that a term like $\Tr ( \gamma_5\, g^{-1} \del_\nu g)$ reduces to the form $\del_\nu \vf$ for the axial $U(1)$ case
of $g = e^{i \gamma_5 \vf}$.)
In Maxwell's equations, the effect of the extra term
(\ref{gur19}) is to give an electromagnetic current
\beq
J^\mu = {e^2 \over 2 \pi^2} \epsilon^{\mu\nu\alpha\beta}
V_\nu\, \del_\alpha A_\beta
\label{gur20}
\eeq
Further, in an equilibrium plasma of quarks and gluons, ${\dot \vf}$
can be replaced by a difference of the chemical potentials
for the left and right chiral fermion numbers as
${\dot \vf} \rightarrow {1\over 2} (\mu_L - \mu_R)$, so that
(\ref{gur20}) gives
\beq
{\vec J}  = - {e^2 \over 4 \, \pi^2} \, (\mu_L - \mu_R) \, {\vec B}
\label{gur21}
\eeq
This shows that the chiral asymmetry in the quark-gluon plasma leads to an electromagnetic current along the direction of the magnetic field.
This is the chiral magnetic effect; it was originally obtained via
Feynman diagram calculations \cite{CME1}, but is very straightforward once we think of fluids in terms of groups. 

Just for the sake of completeness, since the chiral magnetic effect is such a novel twist on the chiral anomaly, let me make a few remarks on the
experimental evidence for it in heavy ion collisions as carried out at
the Relativistic Heavy Ion Collider (RHIC) at Brookhaven
or the Large Hadron Collider (LHC) at CERN, Geneva.
The slightly off-center collision of two large nuclei can create a small blob
of the quark-gluon plasma, because the compression effect from the collision with a large amount of energy being squeezed into a small volume creates
a transition from the confined hadronic phase to the plasma phase
(see Fig.\,{\ref{fig5}}).
The two nuclei flying by, since they are electrically charged, are equivalent to electric currents and produce a magnetic field perpendicular to the collision plane. The field can be as large as $10^{17}$ gauss, which is very high although it lasts
only for a very short time. But this time-scale can be long enough on the scale of strong nuclear interactions that
 the small blob of plasma in the center can be considered as a near-equilibrium
state with a magnetic field. Equation (\ref{gur21}) should apply,
and should be observable as a charge asymmetry between
the final-state particles coming off above the reaction plane and below it.
(The value of $\mu_L - \mu_R$
can be estimated in terms of estimate of the number of instantons
in the plasma.)

\begin{figure}[!t]
\begin{center}
\scalebox{.3}{\includegraphics{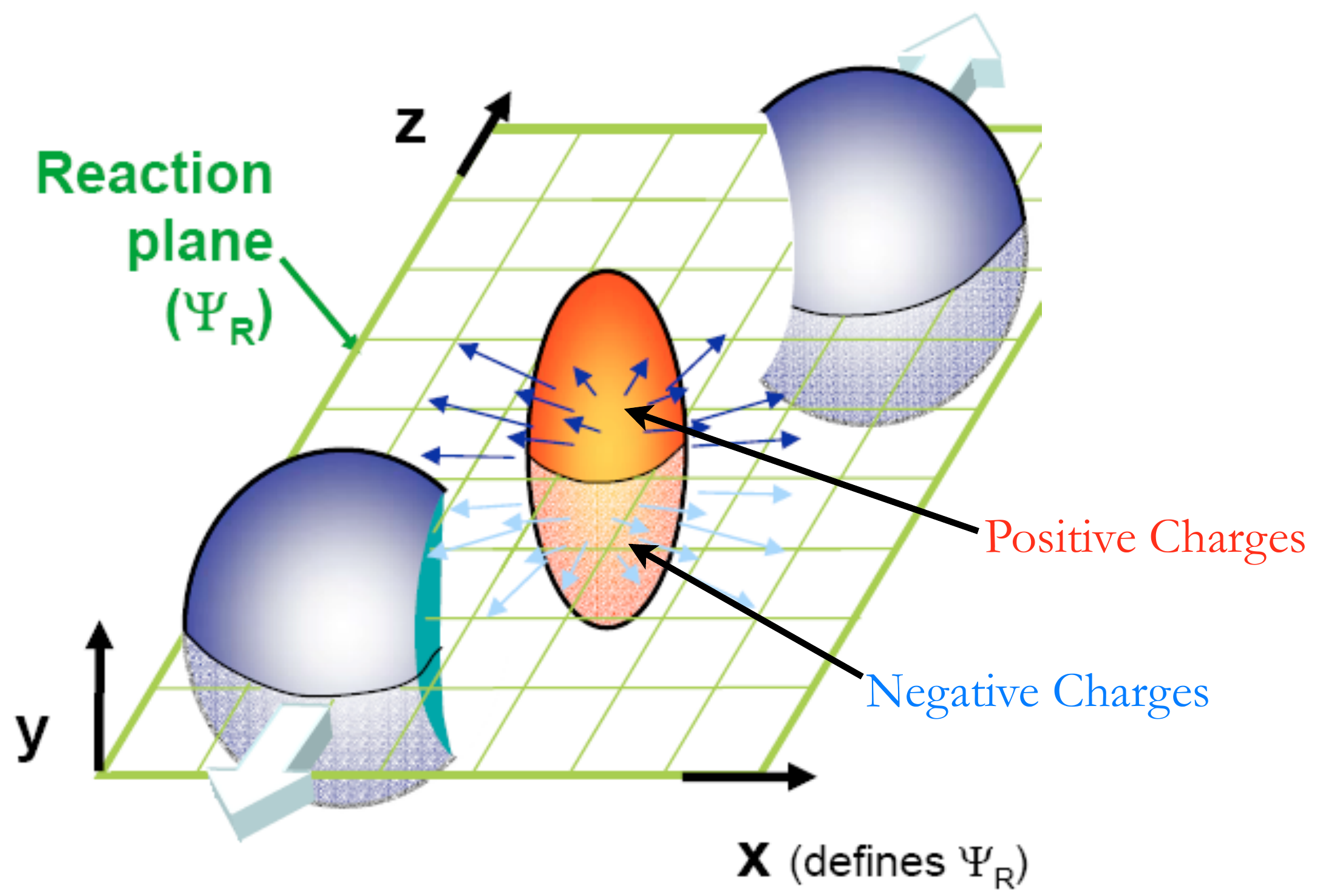}}\hskip .1in
\scalebox{.7}{\includegraphics{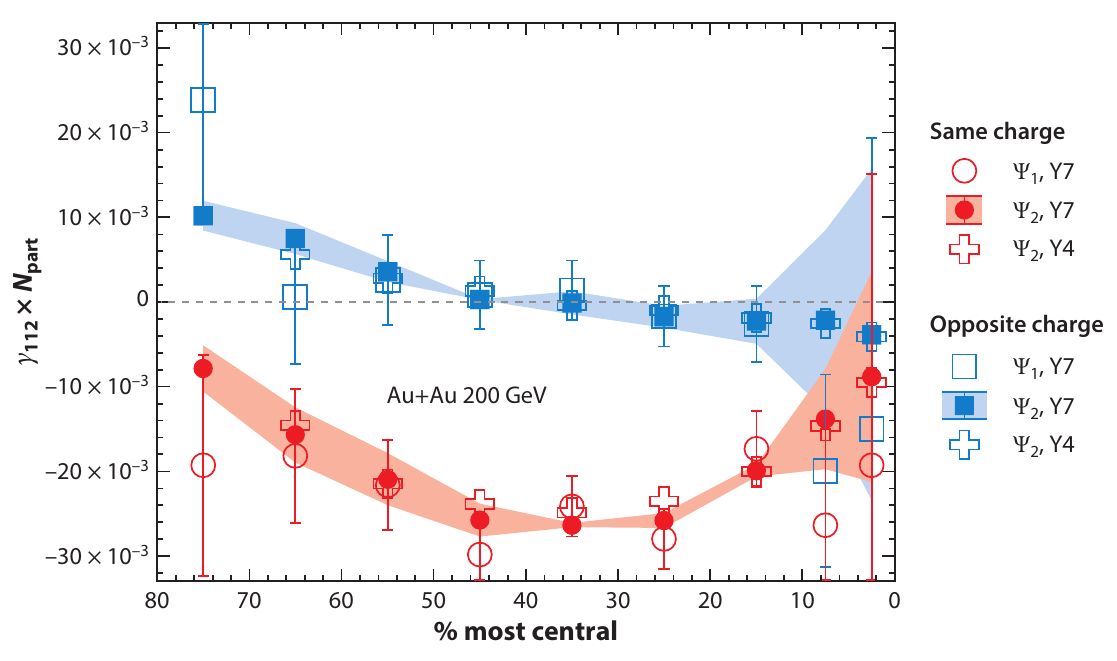}}
\end{center}
\caption{Schematics of the heavy ion collision and the data on $\gamma_{112}$ versus centrality.}
\label{fig5}
\end{figure}

The experimental signal one uses to search for this involves a three-particle correlation, focusing on two particles (of similar or opposite charge)
with a third used to identify the reaction plane \cite{CME2}.
This correlator, denoted usually by $\gamma_{112}$,
will have different values, event-by-event, for the reference particles
having the same sign of charge (SS) or opposite charge (OS).
A plot of the aggregated results over all events in the analysis
is shown in
Fig.\,{\ref{fig5}.
The difference $\gamma_{OS} - \gamma_{SS}$ eliminates many
common contributions and zeroes in on the parity-violating part, so this is the
quantity to focus on for chiral asymmetric effects. 
We expect this difference to vanish for central collisions
(since the produced magnetic field should be much smaller)
and for peripheral collisions (which do not produce enough of the plasma
state), i.e., at the ends of the centrality axis.
The difference in the middle of the axis is a clear indication of
the parity asymmetry, but the identification of this with the
chiral magnetic effect (\ref{gur21}) is not conclusive. There are other sources for the charge asymmetry, which are difficult to estimate and subtract out.
(The decay of the $\rho^0$-meson is one example, there are other sources as well.)
 But a fraction of the observed effect, presumably, can be attributed to
(\ref{gur21}). For more details, see \cite{CME2}.

After this excursion into fluid dynamics, I hope it is clear that, 
following the philosophy of looking at problems, even old ones,
in terms of groups and symmetry, can give new insights. Sometimes we are
lucky and we can even get new results.
Bringing beautiful mathematical structures to bear on real physical
problems was the tenor of Feza's life and work. 
This little vignette of fluid dynamics
in terms of group theory is, hopefully, an appropriate
token of appreciation, as we celebrate his centenary year.

\vskip .2in

I thank Professor Seckin K\"urk\c{c}\"uo\v{g}lu and the Physics Department of
the Middle East Technical University in Ankara for
the invitation to speak in commemoration of Feza G\"ursey's 
birth centenary. I also thank the members of the audience, many of
whom made comments and shared 
reminiscences which contributed to a lively
and well-informed celebration.

This work was supported in part by the U.S. National Science Foundation Grants No. PHY-2112729 and No. PHY-1820271.



\begin{thebibliography}{99}
\bibitem{GTze} F. G\"ursey and Hsiung Chia Tze, Ann. of Phys.
{\bf 128}, 29 (1980).

\bibitem{sud} A. Sudbery, Math. Proc. Camb. Phil. Soc.
{\bf 85}, 199 (1979).

\bibitem{EGO} M. Evans, F. G\"ursey and V. Ogievetsky, \PR~{\bf D47},
3496 (1993).

\bibitem{NS} V.P. Nair and J. Schiff, \PL~{\bf 246B}, 423 (1990);
\NP~{\bf B371}, 329 (1992).

\bibitem{Gur1} F. G\"ursey, Il Nuovo Cimento {\bf 16}, 230 (1960).

\bibitem{GT} M.L. Goldberger and S. Treiman, \PR~ {\bf 110}, 1178
(1958).

\bibitem{currenta} For a general reference on current algebra, see
S.B. Treiman, R. Jackiw and D.J. Gross, {\it Lectures on Current
Algebra and its Applications}, Princeton University
Press (1972); 
S.B. Treiman, R. Jackiw, B. Zumino and E. Witten, 
{\it Current Algebra and Anomalies}, World Scientific
Publishing Co. (1985).

\bibitem{AW} S. Adler, \PRL~{\bf 14}, 1051 (1965);
W.I. Weisberger, \PRL~{\bf 14}, 1047 (1965).

\bibitem{G-ML} M. Gell-Mann and M. Levy, Il Nuovo Cimento {\bf 16}, 705
(1960).

\bibitem{Wein1} S. Weinberg, \PRL~{\bf 17}, 616 (1966).

\bibitem{BGN} A.P. Balachandran, M.G. Gundzik and F. Nicodemi,
Il Nuovo Cimento {\bf 44 A}, 1257 (1966).

\bibitem{Tom} Y. Tomozawa, Il Nuovo Cimento {\bf 46 A}, 707 (1966).

\bibitem{RSud} K. Raman and E.C.G. Sudarshan, 
\PL~{21}, 450 (1966); \PR~{\bf 154}, 1499 (1967).

\bibitem{Wein2} S. Weinberg, \PRL~{\bf 18}, 188 (1967).

\bibitem{CGur} P. Chang and F. G\"ursey, \PR~{\bf 164}, 1752 (1967).

\bibitem{GRad} F. G\"ursey and L.A. Radicati, \PRL~{\bf 13},
173 (1964).

\bibitem{Sak} B. Sakita, \PR~{\bf 136}, B1756 (1964).

\bibitem{GRS} F. G\"ursey, P. Ramond and P. Sikivie,
\PL~{60B}, 177 (1976).

\bibitem{GCat} S. Catto and F. G\"ursey, 
 Il Nuovo Cimento {\bf 86 A}, 201 (1985);
 {\it ibid.} {\bf 99 A}, 685 (1988).

\bibitem{Iach} F. Iachello  1980 \PRL~{\bf 44}, 772 (1980);
F. Iachello and A. Arima  {\it The interacting boson model}, Cambridge University Press, 1987; 
H. Miyazawa, \PR~{\bf 170}, 1586 (1968).

\bibitem{Fl-rev} For a review of
fluid dynamics from a modern point of view, see
R. Jackiw, V.P. Nair, S-Y. Pi and
A.P. Polychronakos, J. Phys. A: Math. Gen. {\bf 37}, R327 (2004).

\bibitem{geom} These results are well reviewed in the context of geometric
quantization; some general references, from which the original articles can be traced, are:\\
N.M.J. Woodhouse, {\it Geometric Quantization}, Clarendon Press (1992); 
J. Sniatycki, {\it Geometric Quantization and Quantum Mechanics}, Springer-Verlag (1980);
S.T. Ali and M. Englis,
{\it Quantization Methods: A Guide for Physicists and Analysts}, arXiv:math-ph/0405065;
V.P. Nair, Elements of Geometric Quantization and
Applications to Particles and Fluids,
arXiv:1606.06407[hep-th];
M. Blau, {\it Symplectic Geometry and geometric quantization},
\verb+http://www.blau.itp.unibe.ch/Lecturenotes.html+\\
P. Woit, {\it Quantum Theory, Groups and Representations: An Introduction},\\
\verb+http://www.math.columbia.edu/\%7Ewoit/QM/qmbook.pdf+

\bibitem{Wong} S. K. Wong, {Il Nuovo Cim.} {\bf 65 A}, 689 (1970).

\bibitem{Bal+} A.P. Balachandran, G. Marmo and A. Stern, Nucl. Phys.
B162, 385 (1980);
A.P. Balachandran, G. Marmo, A. Stern and B.S. Skagerstam,
Phys. Lett. {\bf 89B}, 1991 (1980); 
A. P. Balachandran, G. Marmo,
B-S. Skagerstam and A. Stern, \textit{Gauge Symmetries and Fibre
Bundles}, Lecture Notes in Physics 188 (Springer-Verlag, Berlin,
1982).

\bibitem{Bistro} B. Bistrovic, R. Jackiw, H. Li, V. P. Nair, and S. Y. Pi, Phys. Rev.~{\bf D67}, 025013 (2003).

\bibitem{NRR} V.P. Nair, Rashmi Ray and Shubho Roy, 
Phys. Rev. {\bf D86}, 025012(2012).

\bibitem{KarN} D. Karabali and V.P. Nair, Phys. Rev. {\bf D 90}, 105018 (2014).

\bibitem{CNT} D. Capasso, V.P. Nair
and J. Tekel, Phys. Rev. {\bf D88}, 085025 (2013).

\bibitem{MathP} M.~Mathisson, Acta Phys.\ Polon.\  {\bf 6}, 163 (1937). A.~Papapetrou, Proc.\ Roy.\ Soc.\ Lond.\  A {\bf 209}, 248 (1951).

\bibitem{MAN} G. Monteiro, A. Abanov and V.P. Nair, Phys. Rev. {\bf D 91}, 125033
(2015).

\bibitem{CME1} 
  D.~Kharzeev,
  Phys.\ Lett.\  B {\bf 633}, 260 (2006)
  [arXiv:hep-ph/0406125];
  D.~Kharzeev and A.~Zhitnitsky,
  Nucl.\ Phys.\ A\ {\bf 797}, 67  (2007)
  [arXiv:0706.1026 [hep-ph]];
  D.~E.~Kharzeev, L.~D.~McLerran and H.~J.~Warringa,
  Nucl.\ Phys.\ A\ {\bf 803}, 227  (2008)
  [arXiv:0711.0950 [hep-ph]];
  K.~Fukushima, D.~E.~Kharzeev and H.~J.~Warringa,
  Phys.\ Rev.\ D\ {\bf 78}, 074033  (2008)
  [arXiv:0808.3382 [hep-ph]];
  D.~E.~Kharzeev,
  Annals Phys.\ \ {\bf 325}, 205  (2010)
  [arXiv:0911.3715 [hep-ph]].


\bibitem{CME2} STAR Collaboration,
\PRL~{\bf 103}, 251601 (2009). For the three-particle correlator, see
S.A. Voloshin, \PR~{\bf C70}, 057901 (2004);
for recent updates (from which the displayed data is taken),
see J. Zhiao and F. Wang, arXiv:1906.11413[nucl-ex];
W. Li and G. Wang, Ann. Rev. Nucl. Part. Sci.
{\bf 70}, 293 (2020).


\end{thebibliography}
\end{document}